\documentstyle[twoside]{article}

\catcode`\@=11
\long\def\@makefntext#1{
\protect\noindent \hbox to 3.2pt {\hskip-.9pt  
$^{{\eightrm\@thefnmark}}$\hfil}#1\hfill}		

\def\@makefnmark{\hbox to 0pt{$^{\@thefnmark}$\hss}}	
	
\def\ps@myheadings{\let\@mkboth\@gobbletwo
\def\@oddhead{\hbox{}
\rightmark\hfil\eightrm\thepage}   
\def\@oddfoot{}\def\@evenhead{\eightrm\thepage\hfil
\leftmark\hbox{}}\def\@evenfoot{}
\def\sectionmark##1{}\def\subsectionmark##1{}}

\oddsidemargin=\evensidemargin
\newcounter{sectionc}\newcounter{subsectionc}\newcounter{subsubsectionc}
\renewcommand{\section}[1] {\vspace{12pt}\addtocounter{sectionc}{1} 
\setcounter{subsectionc}{0}\setcounter{subsubsectionc}{0}\noindent 
	{\tenbf\thesectionc. #1}\par\vspace{5pt}}
\renewcommand{\subsection}[1] {\vspace{12pt}\addtocounter{subsectionc}{1} 
	\setcounter{subsubsectionc}{0}\noindent 
	{\bf\thesectionc.\thesubsectionc. {\kern1pt \bfit #1}}\par\vspace{5pt}}
\renewcommand{\subsubsection}[1] {\vspace{12pt}\addtocounter{subsubsectionc}{1}
	\noindent{\tenrm\thesectionc.\thesubsectionc.\thesubsubsectionc.
	{\kern1pt \tenit #1}}\par\vspace{5pt}}
\newcommand{\nonumsection}[1] {\vspace{12pt}\noindent{\tenbf #1}
	\par\vspace{5pt}}

\newcounter{appendixc}
\newcounter{subappendixc}[appendixc]
\newcounter{subsubappendixc}[subappendixc]
\renewcommand{\thesubappendixc}{\Alph{appendixc}.\arabic{subappendixc}}
\renewcommand{\thesubsubappendixc}
	{\Alph{appendixc}.\arabic{subappendixc}.\arabic{subsubappendixc}}

\renewcommand{\appendix}[1] {\vspace{12pt}
        \refstepcounter{appendixc}
        \setcounter{figure}{0}
        \setcounter{table}{0}
        \setcounter{lemma}{0}
        \setcounter{theorem}{0}
        \setcounter{corollary}{0}
        \setcounter{definition}{0}
        \setcounter{equation}{0}
        \renewcommand{\thefigure}{\Alph{appendixc}.\arabic{figure}}
        \renewcommand{\thetable}{\Alph{appendixc}.\arabic{table}}
        \renewcommand{\theappendixc}{\Alph{appendixc}}
        \renewcommand{\thelemma}{\Alph{appendixc}.\arabic{lemma}}
        \renewcommand{\thetheorem}{\Alph{appendixc}.\arabic{theorem}}
        \renewcommand{\thedefinition}{\Alph{appendixc}.\arabic{definition}}
        \renewcommand{\thecorollary}{\Alph{appendixc}.\arabic{corollary}}
        \renewcommand{\theequation}{\Alph{appendixc}.\arabic{equation}}
        \noindent{\tenbf Appendix \theappendixc #1}\par\vspace{5pt}}
\newcommand{\subappendix}[1] {\vspace{12pt}
        \refstepcounter{subappendixc}
        \noindent{\bf Appendix \thesubappendixc. {\kern1pt \bfit #1}}
	\par\vspace{5pt}}
\newcommand{\subsubappendix}[1] {\vspace{12pt}
        \refstepcounter{subsubappendixc}
        \noindent{\rm Appendix \thesubsubappendixc. {\kern1pt \tenit #1}}
	\par\vspace{5pt}}

\newcommand{\textlineskip}{\baselineskip=13pt}
\newcommand{\smalllineskip}{\baselineskip=10pt}

\def\eightcirc{
\begin{picture}(0,0)
\put(4.4,1.8){\circle{6.5}}
\end{picture}}
\def\eightcopyright{\eightcirc\kern2.7pt\hbox{\eightrm c}} 

\newcommand{\copyrightheading}[1]
	{\vspace*{-2.5cm}\smalllineskip{\flushleft
	{\footnotesize International Journal of Modern Physics A, #1}\\
	{\footnotesize $\eightcopyright$\, World Scientific Publishing
	 Company}\\
	 }}

\def\abstracts#1#2#3{{
	\centering{\begin{minipage}{4.5in}\baselineskip=10pt\footnotesize
	\parindent=0pt #1\par 
	\parindent=15pt #2\par
	\parindent=15pt #3
	\end{minipage}}\par}} 

\newcommand{\bibit}{\nineit}

\renewenvironment{thebibliography}[1]
	{\frenchspacing
	 \ninerm\baselineskip=11pt
	 \begin{list}{\arabic{enumi}.}
	{\usecounter{enumi}\setlength{\parsep}{0pt}
	 \setlength{\leftmargin 12.7pt}{\rightmargin 0pt} 
	 \setlength{\itemsep}{0pt} \settowidth
	{\labelwidth}{#1.}\sloppy}}{\end{list}}

\newcounter{itemlistc}
\newcounter{romanlistc}
\newcounter{alphlistc}
\newcounter{arabiclistc}

\newcommand{\fcaption}[1]{
        \refstepcounter{figure}
        \setbox\@tempboxa = \hbox{\footnotesize Fig.~\thefigure. #1}
        \ifdim \wd\@tempboxa > 5in
           {\begin{center}
        \parbox{5in}{\footnotesize\smalllineskip Fig.~\thefigure. #1}
            \end{center}}
        \else
             {\begin{center}
             {\footnotesize Fig.~\thefigure. #1}
              \end{center}}
        \fi}

\newcommand{\tcaption}[1]{
        \refstepcounter{table}
        \setbox\@tempboxa = \hbox{\footnotesize Table~\thetable. #1}
        \ifdim \wd\@tempboxa > 5in
           {\begin{center}
        \parbox{5in}{\footnotesize\smalllineskip Table~\thetable. #1}
            \end{center}}
        \else
             {\begin{center}
             {\footnotesize Table~\thetable. #1}
              \end{center}}
        \fi}

\def\@citex[#1]#2{\if@filesw\immediate\write\@auxout
	{\string\citation{#2}}\fi
\def\@citea{}\@cite{\@for\@citeb:=#2\do
	{\@citea\def\@citea{,}\@ifundefined
	{b@\@citeb}{{\bf ?}\@warning
	{Citation `\@citeb' on page \thepage \space undefined}}
	{\csname b@\@citeb\endcsname}}}{#1}}

\newif\if@cghi
\def\cite{\@cghitrue\@ifnextchar [{\@tempswatrue
	\@citex}{\@tempswafalse\@citex[]}}
\def\citelow{\@cghifalse\@ifnextchar [{\@tempswatrue
	\@citex}{\@tempswafalse\@citex[]}}
\def\@cite#1#2{{$\null^{#1}$\if@tempswa\typeout
	{IJCGA warning: optional citation argument 
	ignored: `#2'} \fi}}

\def\pmb#1{\setbox0=\hbox{#1}
	\kern-.025em\copy0\kern-\wd0
	\kern.05em\copy0\kern-\wd0
	\kern-.025em\raise.0433em\box0}

\def\fnt#1#2{\footnotetext{\kern-.3em
	{$^{\mbox{\scriptsize #1}}$}{#2}}}

\def\fpage#1{\begingroup
\voffset=.3in
\thispagestyle{empty}\begin{table}[b]\centerline{\footnotesize #1}
	\end{table}\endgroup}

\def\runninghead#1#2{\pagestyle{myheadings}
\markboth{{\protect\footnotesize\it{\quad #1}}\hfill}
{\hfill{\protect\footnotesize\it{#2\quad}}}}
\font\tenrm=cmr10
\font\tenit=cmti10 
\font\tenbf=cmbx10
\font\bfit=cmbxti10 at 10pt
\font\ninerm=cmr9
\font\nineit=cmti9

\font\eightrm=cmr8

\def\qed{\hbox{${\vcenter{\vbox{			
   \hrule height 0.4pt\hbox{\vrule width 0.4pt height 6pt
   \kern5pt\vrule width 0.4pt}\hrule height 0.4pt}}}$}}

\begin{document}

\runninghead{The Dispersive Approach to
Electroweak Processes $\ldots$} {The Dispersive Approach to
Electroweak Processes $\ldots$}

\normalsize\textlineskip
\thispagestyle{empty}
\setcounter{page}{1}

\copyrightheading{}			

\vspace*{0.88truein}

\fpage{1}
\centerline{\bf THE DISPERSIVE APPROACH TO ELECTROWEAK PROCESSES}
\vspace*{0.035truein}
\centerline{\bf IN THE BACKGROUND MAGNETIC FIELD
}
\vspace*{0.37truein}
\centerline{\footnotesize GUEY-LIN LIN
}
\vspace*{0.015truein}
\centerline{\footnotesize\it Institute of Physics, National Chiao-Tung University, 1001 Ta-Hsueh
Rd}
\baselineskip=10pt
\centerline{\footnotesize\it Hsinchu 300, Taiwan
}
\vspace*{10pt}

\vspace*{0.21truein}
\abstracts{We propose a new 
method to compute amplitudes of electroweak 
processes in the strong background magnetic field, using 
$\gamma\to e^+e^-$ as an example.
We show that the {\it moments} of $\gamma\to e^+e^-$
width are proportional to the derivatives of photon polarization 
function at the zero energy.  
Hence, the 
pair-production width can be easily calculated 
from the latter by the inverse 
Mellin transform. The prospects of our approach are commented.     }{}{}
\medskip

The electroweak phenomena associated with an intensive background magnetic 
field are rather rich. 
Under a background magnetic field, 
a physical photon can 
decay into an $e^+e^-$ pair or split into two photons. 
Such processes are
relevant to the attenuation of gamma-rays from pulsars\cite{STU,BH}. 
Similarly, 
with a sufficient energy, a neutrino can go through the decays $\nu\to
\nu e^+e^-$\cite{BKM} and  $\nu\to \nu\gamma$\cite{IR}. For processes 
without charged 
fermions in the 
final state, such as $\gamma\to \gamma\gamma$ or 
$\nu\to \nu\gamma$, their decay
widths can be expressed as asymptotic series in $B$ for $B< B_c$. 
However, such 
asymptotic expansions are
not possible for  $\nu\to
\nu e^+e^-$ or $\gamma\to e^+e^-$ since the wave functions of 
final-state fermions are 
non-analytic with respect to the magnetic field strength at $B=0$.

Previously, the photon absorption coefficients due to $\gamma\to e^+e^-$ 
were computed in two different ways.
One either directly squares the $\gamma\to e^+e^-$ amplitude using the exact 
electron (positron) 
wave function in the background magnetic field\cite{toll}, or applies the optical 
theorem on the photon
polarization function $\Pi_{\mu\nu}$\cite{TE}.  In both approaches, 
the results 
are valid only
for $B< B_c$ and $\omega\sin\theta \gg 2m_e$, where $\omega$ is 
the photon energy and
$\theta$ is the angle 
between the magnetic-field direction and the direction of photon propagation. 
It has been pointed out\cite{DH}
that a correct description of $\gamma\to e^+e^-$ near the pair-production 
threshold $\omega \sin\theta \approx 2m_e$
is crucial for astrophysical applications. 
A numerical study taking into account 
the threshold behavior of $\gamma\to e^+e^-$ was also carried out\cite{DH}. 
In this work, we re-examine the previous analytic approaches to 
$\gamma\to e^+e^-$
\cite{toll,TE}, clarifying their implicit assumptions which lead to 
incorrect threshold behavior 
for the above decay. 

Let us follow Ref.~\cite{TE} which begins with the 
proper-time representation\cite{SCH,TS} 
of photon 
polarization function $\Pi_{\mu\nu}$ in the background 
magnetic field:
\begin{eqnarray}
\Pi_{\mu\nu}(q)&=&-{e^3B\over (4\pi)^2}\int_0^{\infty}ds
\int_{-1}^{+1} dv \{e^{-is\phi_0}[(q^2g_{\mu\nu}-
q_{\mu}q_{\nu})N_0 \nonumber \\
&-&(q_{\parallel}^2g_{\parallel\mu\nu}-
q_{\parallel\mu}q_{\parallel\nu})N_{\parallel} 
+(q_{\bot}^2g_{\bot\mu\nu}-
q_{\bot\mu}q_{\bot\nu})N_{\bot}]\nonumber \\
&-&e^{-ism_e^2}(1-v^2)(q^2g_{\mu\nu}-
q_{\mu}q_{\nu})\},
\label{proper_t}
\end{eqnarray}
where
\begin{equation}
\phi_0=m_e^2-{1-v^2\over 4}q_{\parallel}^2-{\cos(zv)-\cos(z)\over 2z\sin(z)}
q_{\bot}^2,
\end{equation}
with $z=eBs$, and $N_{0,\parallel,\bot}$ trigonometric functions of
$z$ and $v$. Here $\parallel$ and $\bot$ are defined 
relative to the magnetic-field
direction.

The photon dispersion relation is given by
$q^2+\rm{Re}\Pi_{\parallel,\bot}=0$,
where $\Pi_{\parallel,\bot}=\epsilon^{\mu}_{\parallel,\bot}
\Pi_{\mu\nu}\epsilon^{\nu}_{\parallel,\bot}$ with 
$\epsilon^{\mu}_{\parallel}$ and $\epsilon^{\mu}_{\bot}$
respectively the the photon polarization vectors parallel and perpendicular
to the plane spanned by the photon momentum ${\bf q}$ and the magnetic field 
${\bf B}$. The imaginary part of $\Pi_{\parallel,\bot}$
is related to photon absorption coefficients 
$\kappa_{\parallel,\bot}$ (i.e. the width of 
$\gamma\to e^+e^-$) via $\kappa_{\parallel,\bot}=
\rm{Im}\Pi_{\parallel,\bot}/\omega$, with $\omega$ the photon
energy. The authors of Ref.~\cite{TE}
analyzed the functions $\Pi_{\parallel,\bot}$ 
in the limit $\omega\sin\theta
\gg 2m_e$ and $B< B_c$. 
They found
$\kappa_{\parallel,\bot}={1\over 2}\alpha\sin\theta 
(eB/m_e)
T_{\parallel,\bot}(\lambda)$, with
\begin{equation}
T_{\parallel,\bot}(\lambda)={4\sqrt{3}\over \pi \lambda}
\int_0^1 dv (1-v^2)^{-1}\left[(1-{1\over 3}v^2), ({1\over 2}+{1\over 6}v^2)
\right]K_{2/3}\left({4\over \lambda}{1\over 1-v^2}\right),
\label{bessel}
\end{equation}     
where $\lambda={3\over 2}(eB/m_e^2)(\omega/m_e)\sin\theta$ and
$K_{2/3}$ is the modified Bessel function. Compared to the 
numerical 
study\cite{DH}, this result is accurate at the higher energy with 
$\xi\equiv \omega^2\sin^2\theta
B_c/2m_e^2B > 10^3$. However its low energy 
prediction is problematic.
One would expect that $T_{\parallel,\bot}(\lambda)$ 
vanishes for $\omega\sin\theta$ below
the pair production threshold. On the other hand, for $\lambda\ll 1$, 
$T_{\parallel,\bot}\to (3/2)^{1/2}\cdot ({1/2}, 
{1/4})e^{-4/\lambda}$,
which does not behave like a step function.
Furthermore $T_{\parallel,\bot}$ is a smooth function of 
$\lambda$, while in actual situation it should contain infinite many 
sawtooth absorption edges corresponding to higher 
Landau levels reachable by the increasing photon energies. 
These 
discrepancies might have to do with the 
approximation made in Ref.~\cite{TE}
where only the small-$s$ contribution in Eq.~(\ref{proper_t}) 
is taken into account. However, due to the highly oscillatory behavior of 
the integrand, it remains unclear how to evaluate the large-$s$ 
contribution.    

Recently, we have developed a technique to deal with the external-field
problem, using the analytic properties of physical amplitudes\cite{KLT}.
We observe that once the amplitude of a physical process is known in the 
small momentum (energy) regime, its behavior at arbitrary momentum 
(energy) is completely determined by the inverse Mellin transform.
For the current problem, we have\cite{KLT}
\begin{equation}
{1\over n!}\left({d^n\over d(\omega^2)^n}
\Pi_{\parallel,\bot}\right)\Big{\vert}_{\omega^2=0}
={M_{\parallel,\bot}^{1-2n}\over \pi}\int_{0}^{1}dy\cdot y^{n-1}
\cdot \left(\kappa_{\parallel
,\bot}(y) y^{-1/2}\right),
\label{mellin}
\end{equation}
where $y=M^2_{\parallel,\bot}/\omega^2$ 
with $M_{\parallel,\bot}$ the threshold energies of pair 
productions\cite{toll,adl} given by
$M^2_{\parallel}\sin^2\theta=4m_e^2$ and
$M^2_{\bot}\sin^2\theta=m_e^2\left(1+\sqrt{1+2{B/B_c}}
\right)^2$.
One notes that the imaginary part of 
$\Pi_{\parallel,\bot}(\omega^2)$ vanishes for the range 
$0 \le \omega^2 \le M^2_{\parallel,\bot}$. 
This property has been verified in the previous works\cite{toll,adl}. 
Therefore one can effectively set the integration range of 
Eq. (\ref{mellin}) as from $y=0$ to $y=\infty$. 
It is then obvious 
that the derivatives of $\Pi_{\parallel,\bot}$ at the zero 
energy are proportional to the Mellin transform of 
$\kappa_{\parallel,\bot}
\cdot y^{-1/2}\equiv 
\kappa_{\parallel,\bot}\cdot \omega/M_{\parallel,\bot}$.
Once the l.h.s. of Eq. (\ref{mellin}) 
is calculated, the absorption coefficients $\kappa_{\parallel,\bot}$ can
be determined by the inverse Mellin transform. 

The l.h.s of Eq.~(\ref{mellin}) is calculated with a rotation of 
integration contour $s\to -is$, which is permissible only for $\omega$
below the pair-production threshold. By this rotation, the phase factor
$\exp(-is\phi_0)$ in Eq.~(\ref{proper_t}) turns into the more well-behaved 
factor $\exp(-s\tilde{\phi_0})$ where $\tilde{\phi_0}$ is obtained 
from $\phi_0$ by the replacement $z\to -iz$ ($z=eBs$). 
Furthermore, the trigonometric
functions in the integrand $N_{0,\parallel,\bot}$  
also turn into the 
hyperbolic function of $z$ and $v$. Due to the presence of 
$\exp(-s\tilde{\phi_0})$ and the assumption of a sub-critical magnetic field
$B < B_c$, one may obtain a first-approximation for
$(d/d\omega^2)^n
\Pi_{\parallel,\bot}{\vert}_{\omega^2=0}$
by disregarding the large-$s$ 
contribution in the integral of Eq.~(\ref{proper_t}) and its 
derivatives. This amounts to approximating, for example,
$\cosh(z)$ there 
by the power series $1+z^2/2+\cdots$. With this approximation,
we arrive at
\begin{eqnarray}
{1\over n!}\left({d^n\over d(\omega^2)^n}
\Pi_{\parallel,\bot}\right)\Big{\vert}_{\omega^2=0}
&=&{2\alpha m_e^2\over \pi}\left({B^2\sin^2\theta\over 3B_c^2m_e^2}\right)^n
{\Gamma(3n-1)\Gamma^2(2n)\over \Gamma(n)
\Gamma(4n)}\nonumber \\
&\times&\left({6n+1, 3n+1\over 4n+1}\right)+\cdots ,
\label{diff}
\end{eqnarray} 
   
where the neglected terms are suppressed by the factor $(B/B_c)^2$.   
Combining Eqs.~(\ref{mellin}) and (\ref{diff}), one obtains the 
photon absorption coefficients $\kappa_{\parallel,\bot}$ by the inverse 
Mellin transform\cite{KLT}:
\begin{eqnarray}
\kappa_{\parallel}&=&{\alpha m_e^2\over i\pi\omega}
\int_{-i\infty+a}^{+i\infty+a} ds {(\lambda^{'})}^{2s}
{\Gamma(3s)\Gamma^2(2s)\over \Gamma(s)\Gamma(4s)}{1\over 3s-1}
\times {6s+1\over 4s+1},\nonumber \\
\kappa_{\bot}&=&{2\alpha m_e^2\over i\pi\omega}{1\over 1+\sqrt{1+2B/B_c}}
\int_{-i\infty+a}^{+i\infty+a} ds {(\lambda^{''})}^{2s}
{\Gamma(3s)\Gamma^2(2s)\over \Gamma(s)\Gamma(4s)}{1\over 3s-1}
\times {3s+1\over 4s+1},\nonumber \\
\label{abso}
\end{eqnarray}  
where $a$ is any real number greater than $1/3$; while $\lambda^{'}=
(\omega\sin\theta B/ \sqrt{3}m_e B_c)$ and
$\lambda^{''}=\lambda^{'}\cdot (1+\sqrt{1+2B/B_c})/2$.
It can be shown rigorously that\cite{KLT} 
$\kappa_{\parallel,\bot}$ computed in this
way are equivalent to results of Tsai and Erber given in Eq.~(\ref{bessel}), except
on some trivial kinematic factors. 

The work on improving Eq. (\ref{diff}) and consequently the threshold 
behavior of $\kappa_{\parallel,\bot}$ is in progress\cite{KLT2}. We have 
found that the large-$s$ contribution which is 
disregarded in the first-approximation becomes important 
in the higher derivatives of $\Pi_{\parallel,\bot}$. Taking into account this
contribution is crucial to obtain correct threshold behaviors for 
$\kappa_{\parallel,\bot}$.


This work is supported in part by the National Science Council of Taiwan
under the grant number NSC89-2112-M009-035.

\nonumsection{References}


\noindent

\begin{thebibliography}{99}
%
\bibitem{STU}
P. A. Sturrock, {\bibit ApJ} {\bf 164}, 529, 1971.
%
\bibitem{BH}
M. G. Baring and A. K. Harding, {\bibit ApJ} {\bf 482}, 372, 1997.
%
\bibitem{BKM}
A. V. Borisov, A. I. Ternov, and V. Ch. Zhukovsky, {\bibit Phys. Lett. B}
{\bf 318}, 489 (1993); A. V. Kuznetsov and N. V. Mikheev, {\it ibid.}
{\bf 394}, 123 (1997).
%
\bibitem{IR}Ara N. Ioannisian and Georg G. Raffelt, {\bibit Phys. Rev. D}
 {\bf 55},
7038 (1996).
%
\bibitem{toll}
J. S. Toll, Ph.D. thesis, Princeton Univ., 1952 (unpublished).
%
\bibitem{TE}
W.-y. Tsai and T. Erber, {\bibit Phys. Rev. D} {\bf 10}, 492, 1974. 
%
\bibitem{DH}
J. K. Daugherty and A. K. Harding, {\bibit ApJ} {\bf 273}, 761, 1983.
%
\bibitem{SCH}
J. Schwinger, {\bibit Phys. Rev.} {\bf 82}, 664, 1951.
%
\bibitem{TS}
W.-y. Tsai, {\bibit Phys. Rev. D} {\bf 10}, 2699, 1974.
%
%
\bibitem{KLT}
W. F. Kao, G.-L. Lin and J.-J. Tseng, hep-ph/0008240, 
to appear in {\bibit Phys. Lett. B.}
%
\bibitem{adl}
S. L. Adler, {\bibit Ann. Phys.} (N.Y.) 67, 599, 1971.
%
\bibitem{KLT2}
W.-F. Kao, G.-L. Lin and J.-J. Tseng, work in progress.
\end{thebibliography}
\end{document}